\documentclass[a4paper,10pt,final]{article}
\usepackage{amsfonts}
\usepackage{amsmath}
\usepackage{amssymb}
\usepackage[dvips]{graphicx}

\DeclareGraphicsExtensions{.bmp,.png,.pdf,.jpg}

\begin{document}

\title{Quantum credit loans}
\author{J. S. Ardenghi$^{\dag }$\thanks{%
email:\ jsardenghi@gmail.com, fax number:\ +54-291-4595142} \\
$^{\dag }$IFISUR, Departamento de F\'{\i}sica (UNS-CONICET)\\
Avenida Alem 1253, Bah\'{\i}a Blanca, Buenos Aires, Argentina}
\maketitle

\begin{abstract}
Quantum models based on the mathematics of quantum mechanics (QM) have been
developed in cognitive sciences, game theory and econophysics. In this work
a generalization of credit loans is introduced by using the vector space
formalism of QM. Operators for the debt, amortization, interest and periodic
installments are defined and its mean values in an arbitrary orthonormal
basis of the vectorial space give the corresponding values at each period of
the loan. Endowing the vector space of dimension $M$, where $M$ is the loan
duration, with a $SO(M)$ symmetry, it is possible to rotate the eigenbasis
to obtain better schedule periodic payments for the borrower, by using the
rotation angles of the $SO(M)$ transformation. Given that a rotation
preserves the length of the vectors, the total amortization, debt and
periodic installments are not changed. For a general description of the
formalism introduced, the loan operator relations are given in terms of a
generalized Heisenberg algebra, where finite dimensional representations are
considered and commutative operators are defined for the specific loan
types. The results obtained are an improvement of the usual financial
instrument of credit because introduce several degrees of freedom through
the rotation angles, which allows to select superposition states of the
corresponding commutative operators that enables the borrower to tune the
periodic installments in order to obtain better benefits without changing
what the lender earns.
\end{abstract}

\section{Introduction}

The mathematical methods of quantum mechanics have spread outside quantum
physics and have reached the field of human condition (\cite{ae1}, \cite{bor}%
, \cite{ae2}, \cite{ae3}, \cite{kh} and \cite{atm}), model of decision
making (\cite{bus}, \cite{ae4}, \cite{mog}, \cite{yuk},\cite{khr}, \cite{pot}%
, \cite{bus2}, \cite{ger} and \cite{kas}), quantum game theory (\cite{mey}, 
\cite{kla}, \cite{eis}, \cite{du} and \cite{mar}), econophysics (\cite{sch}, 
\cite{bag}, \cite{cho}, \cite{ata}, \cite{baq0}, \cite{pio1} and \cite{pio2}%
) among others. This suggests us that the abstract principles of quantum
mechanics exhibit a high degree of adaptability and contextuality and allow
to apply it as "quantum like" models to different branches of human
cognition science. The core of quantum mechanics is the possibility of
superposition and entanglement of quantum states, that physically manifest
as interferences and non-local correlations of observables. Interestingly,
different interpretation of quantum mechanics endow ontological reality to
different mathematical objects of the theory. In particular, modal
interpretations describe the quantum states as the possible properties of
the system and their corresponding probabilities, and not the actual
properties, where properties and facts belong to different ontological
categories (\cite{mod1}, \cite{mod2} and \cite{mod3}). The relationship
between the quantum state and the actual properties of the system is
probabilistic, that is, the actual occurrence of a particular value of an
observable is an indeterministic phenomena. Then, if quantum mechanics
describes the reality, it certainly not an actual reality but a possible
reality. Is the quantum superposition, only existing in potential terms,
that allows to certain measurements to be actualized in a genuinely
unpredictable way. This behavior can be extrapolated to other conceptual
situations, for example in judgments and decisions that can be conceived as
indeterministic processes when subjects give answers in situations of
uncertainty, confusion or ambiguity. In turn, it can be extrapolated to
model stock return distributions by considering the market force as a
quantum harmonic oscillator \cite{ahn} or can be used to apply quantum
probability theory to formulate probabilistic models of cognition \cite{bru}
and to model economical markets \cite{smolin}.

In this sense, the aim of this work is to contribute to the list of possible
extrapolations of the abstract formalism of quantum mechanics to the field
of econophysics. We will show that\ a particular financial instrument, the
credit, can be defined over a vectorial space of dimension $M$, where $M$ is
the duration of the repayment of the credit, which in the moment of
contractual agreement between the lender and the borrower becomes a debt
that is returned in a set of scheduled payments. It will be shown that the
sequence of values for the debt, amortization, interest and periodic
payments can be obtained as the eigenvalues of the respective operators in a
particular eigenbasis of the vectorial space. Then, it will be shown that
the basis chosen of the vector space in which the loan operators are
diagonal can be rotated without changing the total amortization and the
lender's profit, computed as the sum of the periodic payments. This rotation
is a manifestation of a $SO(M)$ symmetry, where $SO(M)$ is the special
orthogonal group of a vectorial space of dimension $M$. Then, by endowing
the vectorial space with a symmetry group, the mean values of the operators
in the rotated basis allow to obtain different configurations for the debt,
amortization and periodic payments, which contain the well known French and
German loan systems, among others. Moreover, it will be shown that the
extrapolation of the abstract mathematical structure of the quantum
formalism to the credit system, can be rewritten in terms of an algebra of
operators, that resembles the Heisenberg algebra of quantum mechanics. In
order to be self-contained, this work will contains some elementary examples
of credit loans and the generalization in a linear space, that we call
\textquotedblleft quantum credit loans\textquotedblright , for the purpose
of introducing the machinery of the basic linear algebra which can be rather
unfamiliar in the context of loans. Then, this manuscript will be organized
as follows:\ In Section II, non-indexed credit loans are reviewed. In
Section III, the recurrence relations for the debt, amortization, interest
and periodic installments are described in terms of a Generalized Heisenberg
algebra. Then the quantum credit loans are introduced, where the rotation
transformation of the orthonormal basis is applied and the two most simple
examples with $M=2$ and $M=3$ are exhaustively studied. In Section IV,
indexed credit loans, where a non-constant interest rate is possible during
the loan, is introduced and a particular example is studied. Finally, the
conclusions are presented.

\section{Non-indexed credit loans}

In order to introduce the algebra of the credit loans, we can notice that
these are a particular case of non-homogeneous linear recurrence relation
for the debt $D$, capitalization or amortization $A$ and interest $Y$.%
\footnote{%
We use $Y$ for interest and not $I$ because $I$ will be used for the
identity operator.} The main difference between the loans lies in the way
the amount of the installments that must be paid are computed in order to
repay these loans. The French system loan is the most popular because the
formula used to calculate periodical fees ensures that the periodic payment
is the same for the life of the loan. This repayment includes the interest
and the repayment of the capital. The only thing that will change is the
proportion of what we pay of these two magnitudes: as the first increases,
the second will decreases, so that the repayments are always the same. Since
the debt $D$ at the beginning of the loan is very large during the initial
periods, the interest $Y$ will also be greater. For this reason, we usually
begin by paying a greater proportion of interest, and as time passes and the
interest payments goes down, we begin to repay a greater proportion of
capital $A$. Another popular system is the German system in which the
periodic payment is not constant but the capital repayment is. Then, in
order to express mathematically the recurrence relations of the French or
German loans, we can consider an initial debt $d_{0}$, a loan duration $M$
and the interest rate $t$. The set of relations between the interest $y_{n}$%
, the debt $d_{n}$ and the amortization $a_{n}$ reads%
\begin{gather}
\text{a) }q_{n}=a_{n}+y_{n}  \label{m1} \\
\text{b) }y_{n}=td_{n-1}  \notag \\
\text{c) }d_{n+1}=d_{n}-a_{n+1}  \notag
\end{gather}%
the first equation \ref{m1}a) means that the periodic payment or installment 
$q_{n}$ is composed of the interest $y_{n}$ and the amortization $a_{n}$ of
the respective periodicity $n$. The second equation \ref{m1}b) implies that
the interest $y_{n}$ is proportional to the debt $d_{n-1}$ of the previous
period and finally the last equation implies that the new debt $d_{n+1}$ is
the subtraction of the debt from the last period with the current
amortization. The coefficient $t$ is the interest rate. The procedure in
order to compute the loan is the following: an initial debt $d_{0}$ is
considered and the lapse $M$ of the loan is needed. With the initial debt,
the interest $y_{1}$ is computed as $y_{1}=td_{0}$. With this value and $%
q_{1}$, the amortization can be computed as $a_{1}=q_{1}-y_{1}$. Finally,
the new debt $d_{1}=d_{0}-a_{1}$ is obtained and the procedure starts again.
As it is expected, the loan must ends, which implies that $d_{M}=0$. The
boundary conditions $d_{0}=d_{0}$ and $d_{M}=0$ define a relation between
the periodic installments $q_{i}$, $d_{0}$ and $M$ and determine univocally
the loan series $d_{n}$, $a_{n}$, $y_{n}$ and $q_{n}$. In turn, by using eq.(%
\ref{m1})c) and summing in $a_{n}$, we obtain that the total amortization $%
\sum\limits_{n=1}^{M}a_{n}=d_{0}$ must be identical to the initial debt
accordingly. In the French system, $q_{n}$ is fixed and does not depend on $%
n $ and in the German system $a_{n}$ is fixed. Combining the three equations
of eq.(\ref{m1}) we obtain the following recurrence relation for $d_{n}$ 
\begin{equation}
d_{n}=(1+t)d_{n-1}-q_{n}  \label{m2}
\end{equation}%
which has the form of a non-homogeneous recurrence relation of the type $%
t_{n}=\alpha t_{n-1}+\beta _{n}$ with $\alpha =1+t$ and $\beta _{n}=-q_{n}$.
The solution of the recurrence relation of eq.(\ref{m2}) with fixed constant 
$q^{(F)}$ reads%
\begin{equation}
d_{n}^{(F)}=\frac{q^{(F)}}{t}\left[ 1-(1+t)^{n}\right] +(1+t)^{n}d_{0}
\label{m2.1}
\end{equation}%
where the superscript $F$ indicates the French system loan. The boundary
condition $d_{M}=0$ implies that%
\begin{equation}
d_{M}^{(F)}=0=(1+t)^{M}\left[ d_{0}+\frac{q^{(F)}}{t}\right] -\frac{q^{(F)}}{%
t}  \label{m2.2}
\end{equation}%
that can be solved for $q^{(F)}$%
\begin{equation}
q^{(F)}=\frac{d_{0}t(1+t)^{M}}{(1+t)^{M}-1}  \label{m2.3}
\end{equation}%
or can be solved for $M$ or $d_{0}$ in the case the value of $q^{(F)}$ is
fixed arbitrarily\footnote{%
This is useful for precancellations of the loan.}. The general procedure to
obtain $q^{(F)}$ is by using $\sum\limits_{n=1}^{M}a_{n}=d_{0}$ and that $%
a_{n}=(1+t)a_{n-1}$ when $q_{n}=q$. Then writing $a_{1}\sum%
\limits_{n=1}^{M}(1+t)^{n}=d_{0}$, where $a_{1}=q-y_{1}=q-td_{0}$ is not
difficult to show eq.(\ref{m2.3}).\ With this result, the debt $d_{n}^{(F)}$
reads%
\begin{equation}
d_{n}^{(F)}=\frac{d_{0}}{(1+t)^{M}-1}\left[ (1+t)^{M}-(1+t)^{n}\right]
\label{m3}
\end{equation}%
In a similar way the amortization and the interest can be solved%
\begin{equation}
a_{n}^{(F)}=(1+t)^{n}(q^{(F)}-td_{0})\text{ \ \ \ \ \ }%
y_{n+1}^{(F)}=td_{n}^{(F)}  \label{m4}
\end{equation}

In the French system, the total amount of unit currency paid is $\mathbf{Q}%
_{F}=q^{(F)}M=\frac{d_{0}Mt(1+t)^{M}}{(1+t)^{M}-1}$. It can be noted from
eq.(\ref{m2}) that the debt contains a constant term $-q^{(F)}$\ and, in the
case that the sequence of numbers for $d_{n}$\ decrease with~$n$, that is, $%
d_{n}<d_{n-1}$, this indicates that $td_{n-1}<q^{(F)}$\ and in the case that
the largest value of the debt is $d_{0}$, then $t<\frac{q^{(F)}}{d_{0}}$,
which implies that, in the case in which the amount to be paid $q^{(F)}$\ in
each period is not fixed, then the interest rate $t$\ must be constrained if
we expect a decreasing debt curve. This issue is not trivial, because in the
case in which $t>\frac{q^{(F)}}{d_{0}}$\ then $d_{n}>d_{n-1}$\ and the debt
increases with $n$.\footnote{%
Indexed interest rate loans will be considered in the next section. In
particular, we will consider a particular case of indexed loan in a new
variable that is linked to the unit currency through the inflation rate (
for example UVA\ system in Argentina).} When the boundary condition $d_{M}=0$%
\ is chosen, then $q^{(F)}=\frac{d_{0}t(1+t)^{M}}{(1+t)^{M}-1}$\ and the
inequality reads $t>0$\ which is obeyed for any loan with positive interest
rate $t$, that is, the boundary condition $d_{M}=0$\ implies that $%
d_{0}>d_{M}$.

\begin{figure}[tbp]
\centering\includegraphics[width=150mm,height=45mm]{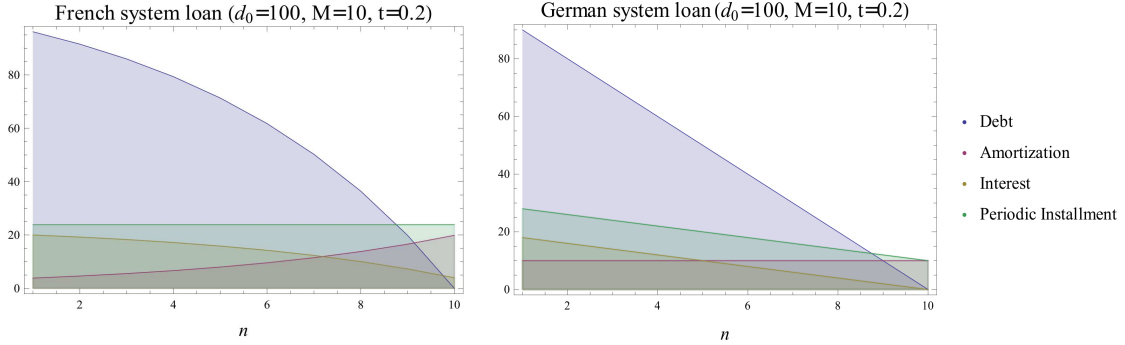}
\caption{Non-indexed credit loan with $d_{0}=100$, $M=10$ and $t=0.2$.
Left:\ French system. Right:\ German system.}
\label{nicl}
\end{figure}

\begin{figure}[tbp]
\centering\includegraphics[width=100mm,height=80mm]{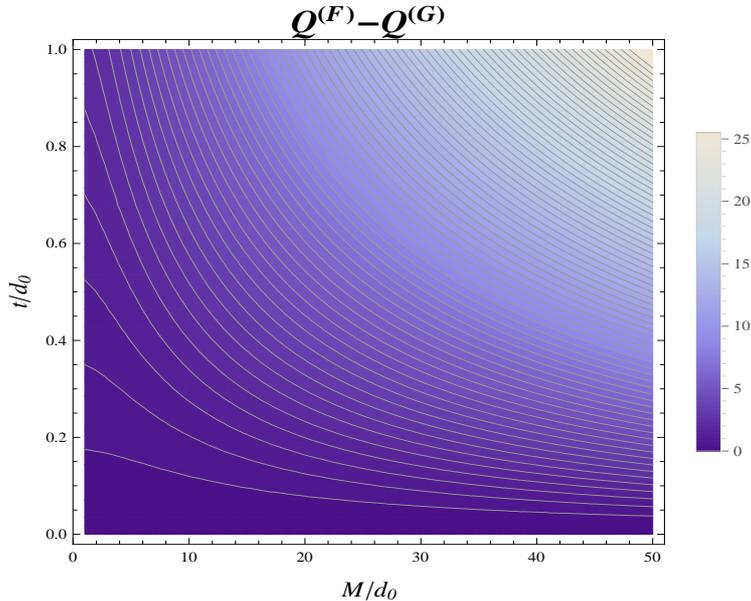}
\caption{$\mathbf{Q}_{F}-\mathbf{Q}_{G}$ difference, where $\mathbf{Q}_{F}$
is the sum of the periodic installments in the French system and $\mathbf{Q}%
_{G}$ is the sum of the periodic installments in the German system for
different values of $t/d_{0}$ and $M/d_{0}$. }
\label{tvsm}
\end{figure}

The German system consists in a constant amortization over time, that is $%
a_{n}=a_{0}=\frac{d_{0}}{M}$\ which is obtained from $\sum%
\limits_{n=1}^{M}a_{n}=d_{0}$. Then the installment to be paid $q_{n}^{(G)}$
is now a function of $n$ due to eq.(\ref{m1}) a) as $q_{n}=y_{n}-a_{0}$.
When the system of recurrence relations is solved we obtain%
\begin{gather}
d_{n}^{(G)}=d_{0}-n\frac{d_{0}}{M}  \label{m5} \\
y_{n}^{(G)}=td_{0}(1-\frac{(n-1)}{M})  \notag \\
q_{n}^{(G)}=\frac{d_{0}}{M}+t\frac{d_{0}}{M}(M-n+1)  \notag
\end{gather}%
The periodic installment is maximum for $n=1$, $q_{1}=d_{0}(t+\frac{1}{M})$,
due to the fact that the interest to be paid at $n=1$ is maximum. In figure %
\ref{nicl}, the debt, the amortization, the interest and the periodic
installment are shown as a function of $n$ for $t=0.02$, $d_{0}=100$ and $%
M=10$ for the French and German systems. As it can be seen, the debt
decreases until it reaches $d_{M}=0$ in both systems and the periodic
installment decreases in the German system. 
The total amount paid during the loan in the German system is $\mathbf{Q}%
_{G}=\overset{M}{\underset{i=1}{\sum }}q_{n}^{(G)}=\frac{d_{0}}{2}\left[
2+t(1+M)\right] $ that can be compared with the obtained from the French
system $\mathbf{Q}_{F}=\overset{M}{\underset{i=1}{\sum }}q_{n}^{(F)}=qM$,
that is larger than $\overset{M}{\underset{i=1}{\sum }}q_{n}^{(G)}$ for
arbitrary values of $M$, $t$ and $d_{0}$ and allows to see why the French
system is more suitable for the lender (see figure \ref{tvsm}). 
The recurrence relations of eq.(\ref{m1}) allow to obtain different schedule
payments $q_{n}$ by simultaneously changing the amortization schedule. In
the French system, $q$ is fixed and in the German system $a_{0}$ is fixed.
Both quantities are related through eq.(\ref{m1}a) which implies that other
configurations for schedule payments should be found, for example in which $Q
$ and $A$ are fixed or vary simultaneously.

\section{Quantum credit loans}

The mathematical structure of quantum mechanics (QM) is constructed over the
concept of Hilbert spaces (see \cite{balle} and \cite{mes}). The energy of a
particle is the Hamiltonian operator that acts on this Hilbert space. The
eigenvalues correspond to the possible energy values of the quantum system.
There is an elegant procedure to obtain the energy eigenvalues of different
physical systems \cite{souza} and \cite{curado}, where a generalized
Heisenberg algebra (GHA) is defined and where the energy eigenvalues can be
obtained from a recurrence relation derived from the algebra. In \cite%
{curado} the linear recurrence relation is studied identical to those
obtained for the debt, amortization and interest. Then, it is interesting to
derive the coupled system of recurrence relations of eq.(\ref{m1}) from an
analogous algebra used in \cite{curado}. Due to the simultaneous magnitudes (%
$D$, $A$, $Y$ and $Q$) with defined values, we can further generalized the
GHA\ by introducing several operators that commute each other as it is done
in \cite{souza} to implement a three-step recurrence relation. In order to
characterize the loan duration, the algebra must be defined over a
finite-dimensional Hilbert space in contrast to the algebras defined in \cite%
{curado}, where an infinite dimensional Hilbert space is constructed. The
procedure to define finite dimensional Harmonic oscillator algebras was
studied in \cite{buzek}. In this work, we will combine both works in order
to construct the finite dimensional vector space of the loan in which the
GHA\ algebra acts. Let us define the following algebra of operators%
\begin{gather}
\text{a) }\left[ D,A\right] =\left[ D,Y\right] =\left[ D,Q\right] =0\text{ \ 
}  \label{new1} \\
\text{b) }aY=f(D)a\text{\ \ \ \ c)}Ya^{\dag }=a^{\dag }f(D)  \notag \\
\text{d) }\left[ D,a\right] =aA~\ \ \ \ \text{e) }\left[ a^{\dag },D\right]
=Aa^{\dag }  \notag \\
\text{f)}\ Q=Y+A\ \ \text{g) }\left[ a,a^{\dag }\right] =A-d_{0}\left\vert
M\right\rangle \left\langle M\right\vert   \notag
\end{gather}%
where $D$ is the debt operator, $A$ is the amortization operator, $Y$ is the
interest operator and $Q$ is the periodic installment operator. We can
consider a set of $M$ orthogonal vectors $\left\vert n\right\rangle $ that
are simultaneous eigenstates of $D$, $A$, $Y$ and $Q$. This is explicit from
eq.(\ref{new1} a) which implies that $D\left\vert n\right\rangle
=d_{n}\left\vert n\right\rangle $, $A\left\vert n\right\rangle
=a_{n}\left\vert n\right\rangle $, $Y\left\vert n\right\rangle
=y_{n}\left\vert n\right\rangle $ and $Q\left\vert n\right\rangle
=q_{n}\left\vert n\right\rangle $, where $d_{n}$, $a_{n}$, $y_{n}$ and $q_{n}
$ are the respective eigenvalues in the eigenbasis that diagonalizes all the
commuting operators. Should be noted that in the French system, the $Q$
operator is degenerate because $q_{n}=q$ and in the German system the
degenerate operator is $A$.\footnote{%
The four operators $D$, $A$, $Y$ and $Q$ are chosen to commute each other
basically because in a classical loan, at each period, the debt, interest,
amortization and periodic installment have a defined value simultaneously,
which resembles the compatibility of observables in quantum mechanics.} The
annihilation operator $a$ and creation operator $a^{\dag }$ acts on the $%
\left\vert n\right\rangle $ basis as%
\begin{eqnarray}
a\left\vert 1\right\rangle  &=&0\text{ \ \ \ \ \ \ \ \ \ \ }a\left\vert
n\right\rangle =N_{n}\left\vert n-1\right\rangle   \label{new2} \\
a^{\dag }\left\vert M\right\rangle  &=&0\text{ \ \ \ \ \ \ \ \ \ }a^{\dag
}\left\vert n\right\rangle =N_{n+1}\left\vert n+1\right\rangle   \notag
\end{eqnarray}%
where $\left\vert M\right\rangle $ is the highest level at which the debt
eigenvalue is\footnote{%
In the quantum harmonic oscillator the ground state is denoted as $%
\left\vert 0\right\rangle $. In this work, the ground state is denoted as $%
\left\vert 1\right\rangle $ that corresponds to $d_{1}=d_{0}-a_{1}$, $%
q_{1}=a_{1}+y_{1}$ and $y_{1}=td_{0}$.}%
\begin{equation}
D\left\vert M\right\rangle =0  \label{new3}
\end{equation}%
that is, the highest debt level eigenvalue is $d_{M}=0$, which implies that
the loan has finished and $M$ is the loan duration and $N_{n}$ are
normalization factors. Eqs.(\ref{new1}) b), d) and e) defines the relation
between the interest, the debt and the amortization. Eq.(\ref{new1}) f) is
the analog to eq.(\ref{m1}) a). Finally, eq.(\ref{new1}) g)\ defines the
commutation relation between $a$ and $a^{\dag }$ in terms of the
amortization operator $A$. Should be noted that $Tr([a,a^{\dagger }])=0$ as
it is expected. All the operators satisfy the Jacobi identity, in particular 
\begin{gather}
\left[ D,\left[ a^{\dag },a\right] \right] +\left[ a,\left[ D,a^{\dag }%
\right] \right] +\left[ a^{\dag },\left[ a,D\right] \right] =0
\label{new5.1} \\
\left[ D,\left[ A,a\right] \right] +\left[ a,\left[ D,A\right] \right] +%
\left[ A,\left[ a,D\right] \right] =0  \notag \\
\left[ D,\left[ Y,a\right] \right] +\left[ a,\left[ D,Y\right] \right] +%
\left[ Y,\left[ a,D\right] \right] =0  \notag
\end{gather}%
that can be obtained by using that $\left[ D,a^{\dag }a\right] =0$ and that $%
\left[ a,Y\right] =aY-Ya=(f(D)-Y)a$. The same result holds by replacing $%
D\rightarrow Q$ in eq.(\ref{new5.1}). In order to understand how the algebra
of eq.(\ref{new1}) works, we can multiply by $a$ to eq.(\ref{new1})f) $%
aQ=aY+aA$ and replacing in eq.(\ref{new1})d) for $aA$ and by using eq.(\ref%
{new1})b) we obtain%
\begin{equation}
\left[ D,a\right] =aQ-f(D)a  \label{new3.1}
\end{equation}%
that applied to a state $\left\vert n\right\rangle $ gives%
\begin{equation}
d_{n}=d_{n-1}+f(d_{n-1})-q_{n}  \label{new3.2}
\end{equation}%
To obtain the credit loan in the French or German system we can replace the
function $f(D)$ with $f(D)=(T-I)D$ where $T$ is the interest rate operator
that satisfies $\left[ T,D\right] =0$, $I$ is the identity operator and $%
T\left\vert n\right\rangle =(t_{n}+1)\left\vert n\right\rangle $. By using
eq.(\ref{new1}) b), d) and f) we obtain%
\begin{equation}
d_{n}=(1+t_{n-1})d_{n-1}-q_{n}~  \label{new4}
\end{equation}%
in the case that $t_{n}=t$ and $q_{n}=q$, last equation is identical to eq.(%
\ref{m2}), where $t$ is the constant interest rate and $q$ is the fixed
installment in the French system. In this case, the $T$ operator is
degenerate in the $\left\vert n\right\rangle $ basis. Using eq.(\ref{new1}%
)g), the creation and annihilation operators can be written in the
eigenbasis of the debt operator as (see eq.(6) of \cite{buzek})%
\begin{equation}
a=\overset{M-1}{\underset{j=1}{\sum }}N_{j+1}\left\vert j\right\rangle
\left\langle j+1\right\vert \text{ \ \ \ \ \ \ \ \ \ }a^{\dag }=\overset{M-1}%
{\underset{j=1}{\sum }}N_{j+1}\left\vert j+1\right\rangle \left\langle
j\right\vert   \label{1.4}
\end{equation}%
By using last equation, the commutator $\left[ a,a^{\dag }\right] $ can be
written as%
\begin{equation}
\left[ a,a^{\dag }\right] =\overset{M-1}{\underset{j=1}{\sum }}%
(N_{j+1}^{2}-N_{j}^{2})\left\vert j\right\rangle \left\langle j\right\vert
-N_{M}^{2}\left\vert M\right\rangle \left\langle M\right\vert   \label{1.5}
\end{equation}%
where we have define $N_{1}=0$. Combining last equation with the l.h.s. of
eq.(\ref{new1}) g) we obtain%
\begin{equation}
N_{n+1}^{2}-N_{n}^{2}=a_{n}\text{ \ \ for }n=1,2,...,M-1\text{ \ \ \ \ \ \ \
\ \ \ }N_{M}^{2}=d_{0}-a_{M}  \label{1.6}
\end{equation}%
and the solution reads%
\begin{equation}
N_{j}^{2}=\overset{j-1}{\underset{i=1}{\sum }}a_{i}  \label{1.7}
\end{equation}%
For instance, consider the case in which $r_{i}=r=1+t$ and $q_{n}=q$ in the
algebra defined in eq.(\ref{new1}), then the non-trivial commutation
relations reduces to%
\begin{gather}
\text{a) \ }Da^{\dag }=a^{\dag }((1+t)D-Q)\text{ \ \ \ b) \ }aD=((1+t)D-Q)a%
\text{\ \ \ \ \ }  \label{1.8} \\
\text{c) }\left[ a,a^{\dag }\right] =A-d_{0}\left\vert M\right\rangle
\left\langle M\right\vert   \notag
\end{gather}%
which defines the desired recurrence relation for the debt in the French
system. A similar analysis can be done with the German system that only
differs from the French system in the choice of the non-degenerate operator,
that in the German system is $A$, in contrast to the French system that
choices $Q$. The function $f(D)$ is identical in both systems.

At this point is interesting to discuss the loan evolution during time. At
each step $n$ of the loan, the debt, the interest, the amortization and the
periodic installment actualizes. This can be represented by the mean values
of these operators in the state $\left\vert n\right\rangle $, that is 
\begin{eqnarray}
\left\langle n\right\vert D\left\vert n\right\rangle  &=&d_{n}\text{ \ \ \ \
\ \ }\left\langle n\right\vert Y\left\vert n\right\rangle =y_{n}  \label{2.0}
\\
\left\langle n\right\vert A\left\vert n\right\rangle  &=&a_{n}\text{ \ \ \ \
\ \ \ }\left\langle n\right\vert Q\left\vert n\right\rangle =q_{n}  \notag
\end{eqnarray}%
In turn, $\left\vert n\right\rangle =\frac{(a^{\dag })^{n}}{%
\prod\limits_{j=1}^{n}N_{j}}\left\vert 1\right\rangle $, then for instance,
the mean value of the debt operator reads%
\begin{equation}
d_{n}=\frac{1}{\prod\limits_{j=1}^{n}N_{j}^{2}}\left\langle 1\right\vert
a^{n}D(a^{\dag })^{n}\left\vert 1\right\rangle \text{ \ }  \label{2.1}
\end{equation}%
and by using eq.(\ref{new1})e), eq.(\ref{new1})f) and eq.(\ref{new1}) c) and
that for the French system $\left[ Q,a\right] =\left[ Q,a^{\dagger }\right]
=0$ is not difficult to show that%
\begin{equation}
d_{n}=\left( \prod\limits_{j=1}^{n}N_{j}^{2}\right) ^{-1}\left\langle
1\right\vert a^{n}(a^{\dag })^{n}\left\vert 1\right\rangle \left[
(1+t)^{n}d_{0}-q\overset{n-1}{\underset{j=0}{\sum }}(1+t)^{j}\right] 
\label{2.2}
\end{equation}%
but $\left( \prod\limits_{j=1}^{n}N_{j}^{2}\right) ^{-1}\left\langle
1\right\vert a^{n}(a^{\dag })^{n}\left\vert 1\right\rangle =1$ and the
result is identical to eq.(\ref{m2.1}). But then we can notice that the
basis $\left\vert n\right\rangle $ allows to construct linear combinations,
which is reasonable because the state $\left\vert n\right\rangle $ is an
arbitrary unit vector in the vector space and it has the same status as any
linear combination $\,\sum\limits_{n=1}^{M}c_{n}\left\vert n\right\rangle $.
A linear combination of states of the loan enlarges the degrees of freedom
and in turn allows the time evolution not to be constrained to the
eigenvalues $d_{n}$, $y_{n}$, $a_{n}$ and $q_{n}$ of each operator at each
periodicity of the loan. Besides this, the superposition of the loan states
can be used to design loans that, in particular, can be a mixture of the
French and the German systems. To be clear, we can consider $M=2$, then the
operators $D$, $A$, $Y$, $Q$, $a$ and $a^{\dagger }$ are written in the $%
\{\left\vert 1\right\rangle ,\left\vert 2\right\rangle \}$ basis as%
\begin{eqnarray}
D &=&\left( 
\begin{array}{cc}
d_{1} & 0 \\ 
0 & 0%
\end{array}%
\right) \text{ \ \ \ }A=\left( 
\begin{array}{cc}
a_{1} & 0 \\ 
0 & a_{2}%
\end{array}%
\right) \text{ \ \ \ }Y=\left( 
\begin{array}{cc}
y_{1} & 0 \\ 
0 & y_{2}%
\end{array}%
\right)   \label{3} \\
Q &=&\left( 
\begin{array}{cc}
q_{1} & 0 \\ 
0 & q_{2}%
\end{array}%
\right) \text{ \ \ \ }a=\left( 
\begin{array}{cc}
0 & N_{2} \\ 
0 & 0%
\end{array}%
\right) \text{ \ \ \ \ \ }a^{\dagger }=\left( 
\begin{array}{cc}
0 & 0 \\ 
N_{2} & 0%
\end{array}%
\right)   \notag
\end{eqnarray}%
For example in the French system we have%
\begin{gather}
D^{(F)}=d_{0}\left( 
\begin{array}{cc}
\frac{(1+t)}{(2+t)} & 0 \\ 
0 & 0%
\end{array}%
\right) \text{ \ \ \ }A^{(F)}=d_{0}\left( 
\begin{array}{cc}
\frac{1}{2+t} & 0 \\ 
0 & \frac{1+t}{2+t}%
\end{array}%
\right) \text{ \ \ \ }Y^{(F)}=d_{0}\left( 
\begin{array}{cc}
t & 0 \\ 
0 & t\frac{1+t}{2+t}%
\end{array}%
\right)   \label{4} \\
Q^{(F)}=d_{0}\left( 
\begin{array}{cc}
\frac{(1+t)^{2}}{2+t} & 0 \\ 
0 & \frac{(1+t)^{2}}{2+t}%
\end{array}%
\right) \text{ \ \ \ }a=\sqrt{d_{0}}\left( 
\begin{array}{cc}
0 & \sqrt{\frac{1}{2+t}} \\ 
0 & 0%
\end{array}%
\right) \text{ \ \ \ \ \ }a^{\dagger }=\sqrt{d_{0}}\left( 
\begin{array}{cc}
0 & 0 \\ 
\sqrt{\frac{1}{2+t}} & 0%
\end{array}%
\right)   \notag
\end{gather}%
and in the German system we have%
\begin{gather}
D^{(G)}=d_{0}\left( 
\begin{array}{cc}
\frac{1}{2} & 0 \\ 
0 & 0%
\end{array}%
\right) \text{ \ \ \ }A^{(G)}=d_{0}\left( 
\begin{array}{cc}
\frac{1}{2} & 0 \\ 
0 & \frac{1}{2}%
\end{array}%
\right) \text{ \ \ \ }Y^{(G)}=d_{0}\left( 
\begin{array}{cc}
t & 0 \\ 
0 & \frac{t}{2}%
\end{array}%
\right)   \label{5} \\
Q^{(G)}=d_{0}\left( 
\begin{array}{cc}
\frac{1}{2}+t & 0 \\ 
0 & \frac{1}{2}+\frac{t}{2}%
\end{array}%
\right) \text{ \ \ \ }a=\sqrt{d_{0}}\left( 
\begin{array}{cc}
0 & \frac{1}{\sqrt{2}} \\ 
0 & 0%
\end{array}%
\right) \text{ \ \ \ \ \ }a^{\dagger }=\sqrt{d_{0}}\left( 
\begin{array}{cc}
0 & 0 \\ 
\frac{1}{\sqrt{2}} & 0%
\end{array}%
\right)   \notag
\end{gather}%
These operators will be used in the next section in order to introduce the
linear combinations. We can note that the condition $\sum%
\limits_{n=1}^{M}a_{n}=d_{0}$ can be written more compactly as $Tr(A)=d_{0}$%
. In turn, the total amount of unit currency paid $\mathbf{Q}$ by the
borrower is $\mathbf{Q}=Tr(Q)$.

Then, in order to generalize the vector space formalism to $M$ dimensions,
let us consider an orthogonal basis of an $M$ dimensional vectorial space
where each orthogonal vector is denoted as $\left\vert n\right\rangle $ and
let us construct $M$ orthogonal linear combinations%
\begin{equation}
\left\vert \varphi _{n}\right\rangle
=\sum\limits_{j=1}^{M}c_{j}^{(n)}\left\vert j\right\rangle \text{ \ \ \ \ \ }%
j=1,2,...,M  \label{6}
\end{equation}%
where $c_{j}^{(n)}$ are the coefficients of the superposition and obey%
\begin{equation}
\sum\limits_{j=1}^{M}(c_{j}^{(m)})^{\ast }c_{j}^{(n)}=\delta _{nm}\text{ for 
}n\neq m  \label{7}
\end{equation}%
the eq.(\ref{6}) can be written compactly as $\left\vert \varphi
\right\rangle =U\left\vert \varphi _{0}\right\rangle $, where $\left\vert
\varphi \right\rangle =\left( \left\vert \varphi _{1}\right\rangle \text{ }%
\left\vert \varphi _{2}\right\rangle \text{ }\cdots \text{ }\left\vert
\varphi _{M}\right\rangle \right) ^{T}$, $\left\vert \varphi
_{0}\right\rangle =\left( \left\vert 1\right\rangle \text{ }\left\vert
2\right\rangle \text{ }\cdots \text{ }\left\vert M\right\rangle \right) ^{T}$
and%
\begin{equation}
U=\left( 
\begin{array}{cccc}
c_{1}^{(1)} & c_{1}^{(2)} & ... & c_{M}^{(1)} \\ 
c_{2}^{(1)} & c_{2}^{(2)} & \cdots & c_{2}^{(M)} \\ 
\vdots & \vdots & \ddots & \vdots \\ 
c_{M}^{(1)} & c_{M}^{(2)} & \cdots & c_{M}^{(M)}%
\end{array}%
\right)  \label{7.1}
\end{equation}%
The similarity transformation $U$ obeys $U^{T}U=I$ which is obtained by
computing the scalar product $\left\langle \varphi \mid \varphi
\right\rangle =\left\langle \varphi _{0}\right\vert U^{T}U\left\vert \varphi
_{0}\right\rangle =1$. The transformation $U$ belongs to the $SO(M)$ group
symmetry that preserves the distance of a vector under rotation. The $SO(M)$
is the special orthogonal group in $M$ dimensions, and the transformation $U$
is represented by a $M\times M$ matrix elements whose columns and rows are
orthogonal unit vectors with $\det U=1$ \cite{geo}. The $SO(M)$ is a Lie
group which can be obtained through its Lie algebra $so(M)$ by using matrix
exponential of the generators of the algebra \cite{geo}. The number of the
generators of the Lie algebra $so(M)$ is $g(M)=M(M-1)/2$ and is the
dimension of the parameter space of $SO(M)$. These parameters are angles,
then $so(M)$ is a compact group. The parameter space dimension of the $SO(M)$
is larger than the loan duration $M$ for $M>3$ which implies that the
rotation transformation in the vector space of dimension $M$ provides a
large number of degrees of freedom in order to tune the payment schedule
with better benefits for the borrower. The transformation $U\in SO(M)$
induces a transformation on any operator $O$ as $\overline{O}=UOU^{-1}$
where $O$ can be $D$, $A$, $Y$, $Q$, $a$ and $a^{\dagger }$. For example, if 
$a\left\vert n\right\rangle =N_{n}\left\vert n-1\right\rangle $, then $%
UaU^{-1}U\left\vert n\right\rangle =N_{n}U\left\vert n-1\right\rangle $ and
by calling $\overline{a}=UaU^{-1}$, $\left\vert \varphi _{n-1}\right\rangle
=U\left\vert n-1\right\rangle $ and $\left\vert \varphi _{n}\right\rangle
=U\left\vert n\right\rangle $, we can write $\overline{a}\left\vert \varphi
_{n}\right\rangle =N_{n}\left\vert \varphi _{n-1}\right\rangle $, where $%
\left\vert \varphi _{n-1}\right\rangle $ and $\left\vert \varphi
_{n}\right\rangle $ are two orthogonal vectors obtained from the original
orthonormal basis by rotation, that is%
\begin{gather}
\overline{a}\left\vert \varphi _{n}\right\rangle =N_{n}\left\vert \varphi
_{n-1}\right\rangle \text{ \ \ \ }\overline{a}^{\dagger }\left\vert \varphi
_{n}\right\rangle =N_{n+1}\left\vert \varphi _{n+1}\right\rangle
\label{7.1.1} \\
a\left\vert \varphi _{1}\right\rangle =0\text{ \ \ \ \ \ \ \ \ \ \ \ \ \ \ \
\ }a^{\dag }\left\vert \varphi _{M}\right\rangle =0\text{ \ \ }  \notag
\end{gather}%
which means that $\overline{a}$ and $\overline{a}^{\dagger }$ acts as
creation and annihilation operators of loan states in the rotated basis. The
loan operators transform as 
\begin{eqnarray}
\overline{D} &=&UDU^{-1}\text{ \ \ \ \ \ }\overline{A}=UAU^{-1}
\label{7.1.1.1} \\
\overline{Y} &=&UYU^{-1}\text{ \ \ \ \ }\overline{Q}=UQU^{-1}  \notag
\end{eqnarray}%
The mean values of the transformed loan operators in the original basis read%
\begin{eqnarray}
\left\langle n\right\vert \overline{D}\left\vert n\right\rangle &=&\overline{%
d}_{n}\text{ \ \ \ \ \ \ \ }\left\langle n\right\vert \overline{Y}\left\vert
n\right\rangle =\overline{y}_{n}  \label{7.1.2} \\
\left\langle n\right\vert \overline{A}\left\vert n\right\rangle &=&\overline{%
a}_{n}\text{ \ \ \ \ \ \ }\left\langle n\right\vert \overline{Q}\left\vert
n\right\rangle =\overline{q}_{n}  \notag
\end{eqnarray}%
and give the expected values of the loan magnitudes at each period. Should
be noted that it is meaningless to compute $\left\langle \varphi
_{n}\right\vert \overline{O}\left\vert \varphi _{n}\right\rangle $ because $%
\left\langle \varphi _{n}\right\vert \overline{O}\left\vert \varphi
_{n}\right\rangle =\left\langle n\right\vert O\left\vert n\right\rangle $
but $\left\langle n\right\vert \overline{O}\left\vert n\right\rangle
=\left\langle \varphi _{n}\right\vert O\left\vert \varphi _{n}\right\rangle $%
. We can study the formalism for the quantum credit loans with the most
simple examples $M=2$ and $M=3$.

\subsection{$M=2$}

As an example we can consider $M=2$, then it is possible to write the
transformation from the $\{\left\vert 1\right\rangle ,\left\vert
2\right\rangle \}$ to the $\{\left\vert \varphi _{1}\right\rangle
,\left\vert \varphi _{2}\right\rangle \}$ basis as%
\begin{equation}
\left\vert \varphi _{1}\right\rangle =\cos \phi \left\vert 1\right\rangle
+\sin \phi \left\vert 2\right\rangle \text{ \ \ \ \ \ \ }\left\vert \varphi
_{2}\right\rangle =-\sin \phi \left\vert 1\right\rangle +\cos \phi
\left\vert 2\right\rangle   \label{8.0}
\end{equation}%
where $\phi \in \lbrack 0,2\pi ]$, $\left\langle \varphi _{1}\mid \varphi
_{1}\right\rangle =\left\langle \varphi _{2}\mid \varphi _{2}\right\rangle =1
$ and $\left\langle \varphi _{1}\mid \varphi _{2}\right\rangle =0$. The
change of basis $\left\{ \left\vert 1\right\rangle ,\left\vert
2\right\rangle \right\} $ to $\left\{ \left\vert \varphi _{1}\right\rangle
,\left\vert \varphi _{2}\right\rangle \right\} $ can be obtained through a
similarity transformation $U$ as%
\begin{equation}
U=\left( 
\begin{array}{cc}
\cos \phi  & \sin \phi  \\ 
-\sin \phi  & \cos \phi 
\end{array}%
\right)   \label{8}
\end{equation}%
The transformation $U$ belongs to the $SO(2)$ group symmetry and can be
considered as a rotation $\phi $ around the $z$ axis. In this case $g(2)=1$
and the Lie algebra $so(2)$ contains a unique generator $L=\left( 
\begin{array}{cc}
0 & i \\ 
-i & 0%
\end{array}%
\right) $ such that the associated parameter is $\phi $ and $U$ of eq.(\ref%
{8}) can be written as $U=e^{i\phi L}$.\footnote{%
Matrix $L$ is the $\sigma _{y}$ Pauli matrix that represents the intrinsic
angular momentum or spin in the $y$ direction of an electron.} The
transformed annihilation operator reads%
\begin{equation}
\overline{a}=UaU^{-1}=N_{2}\left( 
\begin{array}{cc}
-\cos \phi \sin \phi  & \cos ^{2}\phi  \\ 
-\sin ^{2}\phi  & \cos \phi \sin \phi 
\end{array}%
\right)   \label{9}
\end{equation}%
Then, by applying $\overline{a}$ to $\overline{a}\left\vert \varphi
_{2}\right\rangle $ we obtain that $\overline{a}\left\vert \varphi
_{2}\right\rangle =\left\vert \varphi _{1}\right\rangle $, that is, $%
\overline{a}$ acts as an annihilation operator in the new basis. The
remaining operators read%
\begin{gather}
\overline{D}=UDU^{-1}=\left( 
\begin{array}{cc}
\cos ^{2}\phi d_{1} & \sin \phi \cos \phi d_{1} \\ 
\sin \phi \cos \phi d_{1} & \sin ^{2}\phi d_{1}%
\end{array}%
\right) \text{ \ \ \ \ \ }  \label{9.1} \\
\overline{Y}=UYU^{-1}=\left( 
\begin{array}{cc}
\cos ^{2}\phi y_{1}+\sin ^{2}\phi y_{2} & \sin \phi \cos \phi (y_{1}-y_{2})
\\ 
\sin \phi \cos \phi (y_{1}-y_{2}) & \sin ^{2}\phi y_{1}+\cos ^{2}\phi y_{2}%
\end{array}%
\right)   \notag \\
\overline{A}=UAU^{-1}=\left( 
\begin{array}{cc}
\cos ^{2}\phi a_{1}+\sin ^{2}\phi a_{2} & \sin \phi \cos \phi (a_{1}-a_{2})
\\ 
\sin \phi \cos \phi (a_{1}-a_{2}) & \sin ^{2}\phi a_{1}+\cos ^{2}\phi a_{2}%
\end{array}%
\right) \text{ \ \ \ \ \ }  \notag \\
\text{\ \ \ }\overline{Q}=UQU^{-1}=\left( 
\begin{array}{cc}
\cos ^{2}\phi q_{1}+\sin ^{2}\phi q_{2} & \sin \phi \cos \phi (q_{1}-q_{2})
\\ 
\sin \phi \cos \phi (q_{1}-q_{2}) & \sin ^{2}\phi q_{1}+\cos ^{2}\phi q_{2}%
\end{array}%
\right)   \notag
\end{gather}%
As we saw in the last section, the mean values $\left\langle n\right\vert 
\overline{D}\left\vert n\right\rangle $, $\left\langle n\right\vert 
\overline{A}\left\vert n\right\rangle $, $\left\langle n\right\vert 
\overline{Y}\left\vert n\right\rangle $, $\left\langle n\right\vert 
\overline{Q}\left\vert n\right\rangle \,$\ are the actualization values for
the debt, amortization, interest and periodic installment. For example, the
amortizations read%
\begin{eqnarray}
\overline{a}_{1} &=&\left\langle 1\right\vert \overline{A}\left\vert
1\right\rangle =\cos ^{2}\phi a_{1}+\sin ^{2}\phi a_{2}  \label{10} \\
\overline{a}_{2} &=&\left\langle 2\right\vert \overline{A}\left\vert
2\right\rangle =\sin ^{2}\phi a_{1}+\cos ^{2}\phi a_{2}  \notag
\end{eqnarray}%
and the transformed amortization obeys%
\begin{equation}
\overline{a}_{1}+\overline{a}_{2}=a_{1}+a_{2}=d_{0}  \label{11}
\end{equation}%
as it is expected. This last equation can be obtained for any dimension $M$,
because $Tr(\overline{A})=Tr(UAU^{-1})=Tr(A)=d_{0}$. In turn, the borrower
that takes a loan is interested in the periodic installment schedule, that
can be obtained as 
\begin{eqnarray}
\overline{q}_{1} &=&\left\langle 1\right\vert \overline{Q}\left\vert
1\right\rangle =\cos ^{2}\phi q_{1}+\sin ^{2}\phi q_{2}  \label{12} \\
\overline{q}_{2} &=&\left\langle 2\right\vert \overline{Q}\left\vert
2\right\rangle =\sin ^{2}\phi q_{1}+\cos ^{2}\phi q_{2}  \notag
\end{eqnarray}%
This result is interesting because the total amount paid is $\mathbf{Q}=Tr(%
\overline{Q})=Tr(Q)$, so the basis rotation does not change what the lender
earns. In turn, the basis rotation induces two possible superpositions of
the periodic installments, then, by tuning $\phi $, it is possible to
arrange two different payment schedules $\overline{q}_{1}<\overline{q}_{2}$
or $\overline{q}_{1}>\overline{q}_{2}$. The $\phi $ angle can be a parameter
of the loan that the borrower can control. Interestingly, suppose the case
of a borrower taking a loan of two periodic installments $M=2$. Then,
suppose that before the first payment, a particular event does not allow the
borrower to afford the first periodic payment. But, instead of being
classified as a defaulter, the borrower can manipulate $\phi $ in order to
decrease $q_{1}\rightarrow \overline{q}_{1}$ where $\overline{q}_{1}<q_{1}$.
This can be achieved if the borrower chooses $\phi $ near $\pi /2$
decreasing the weight of $q_{1}$ in $\overline{q}_{1}$ (see figure \ref{a1}%
). Once the basis is rotated, the next periodic installment is $\overline{q}%
_{2}$, that increases the weight of $q_{1}$ in $\overline{q}_{2}$, which
means that $\overline{q}_{2}>q_{2}$. In this sense, the credit loan payments
self-regulate when a rotation of the basis is applied, which is contained in
the relation $Tr(Q)=Tr(\overline{Q})$, which implies for $M=2$, that $%
q_{1}+q_{2}=\overline{q}_{1}+\overline{q}_{2}$ and by calling $\Delta
q_{1}=q_{1}-\overline{q}_{1}$ and $\Delta q_{2}=q_{2}-\overline{q}_{2}$, we
have $\Delta q_{1}=-\Delta q_{2}$ and when $q_{1}>\overline{q}_{1}$ then $%
\overline{q}_{2}>q_{2}$. The manipulation of $\phi $ opens the possibility
to the lender to offer new financial instruments for credit with the same
profit but with better benefits for the borrower than the usual amortization
system. Moreover, in the case of indexed interest rate with some exogenous
variable, such as inflation, the basis rotation of the credit loan is
suitable to control the variations of the periodic payment due to the
fluctuations of the macroeconomy.

There is an inherent risk associated to the uncertainty introduced by the
rotation when the borrower cannot afford the periodic payment. When this
happens, a rotation basis is still available for the borrower, but then,
once the basis is rotated, the payment schedule changes by decreasing the
current payment and increasing the future payment. This inherent risk in the
rotation procedure can be quantified by noting that the basis rotation
introduces superposition of the loan states. One way to quantify this
uncertainty is by computing $\Delta \overline{Q}_{n}=\sqrt{\overline{q^{2}}%
_{n}-\overline{q}_{n}^{2}}$ where $\overline{q}_{n}=\left\langle
n\right\vert \overline{Q}\left\vert n\right\rangle $ and $\overline{q^{2}}%
_{n}=\left\langle n\right\vert \overline{Q}^{2}\left\vert n\right\rangle $.
For example, in the case $M=2$ we obtain%
\begin{eqnarray}
\Delta \overline{Q}_{1} &=&\sqrt{(q_{1}-1)q_{1}\cos ^{2}\phi
+(q_{2}-1)q_{2}\sin ^{2}\phi }  \label{12.1} \\
\Delta \overline{Q}_{2} &=&\sqrt{(q_{2}-1)q_{2}\cos ^{2}\phi
+(q_{1}-1)q_{1}\sin ^{2}\phi }  \notag
\end{eqnarray}%
When $\phi =\pi /2$ and $q_{2}<q_{1}$, last equation implies that $\Delta 
\overline{Q}_{1}=\sqrt{(q_{2}-1)q_{2}}$ and $\Delta \overline{Q}_{2}=\sqrt{%
(q_{1}-1)q_{1}}$, then $\Delta \overline{Q}_{2}>\Delta \overline{Q}_{1}$.
That is, by selecting $\phi =\pi /2$, the borrower is paying $q_{2}$ and not 
$q_{1}$ at the beginning of the loan, but the future risk quantified by $%
\Delta \overline{Q}_{2}$ has increased. The fact that the borrower cannot
paid $q_{1}$ implies that there is a potential risk that it will not be able
to pay it again (it should be noted that for $\phi =\pi /2$, $\overline{q}%
_{1}=q_{2}$ and $\overline{q}_{2}=q_{1}$). But the borrower pays $q_{2}$ at
the beginning of the loan and has time to collect $q_{1}$ meanwhile, so the
risk of not paying $q_{1}$ at the second period of the loan must be less
than the initial risk of not paying $q_{1}$ at the beginning of the loan.
For each rotation basis chosen at each period of the loan there is a risk
flow that should be used as a complement to the quantum credit loans in
order to ensure a measure of the risk to be used by the lender when the
borrower applies a rotation of the loan states. In Table 1 it can be seen
the main difference between the usual credit loan and the quantum credit
loan for $M=2$.

\begin{gather*}
\begin{tabular}{|c|c|c|c|c|}
\hline
$M=2$ & \textbf{Debt }$D$ & \textbf{Interest }$Y$ & \textbf{Amortization }$A$
& \textbf{Installment }$Q$ \\ \hline
Usual & $d_{1}$ & $y_{1}$ & $a_{1}$ & $q_{1}$ \\ \cline{2-5}
Credit Loan & $0$ & $y_{2}$ & $a_{2}$ & $q_{2}$ \\ \hline
Quantum & \multicolumn{1}{|l|}{$d_{1}\cos ^{2}\phi $} & \multicolumn{1}{|l|}{%
$\cos ^{2}\phi y_{1}+\sin ^{2}\phi y_{2}$} & \multicolumn{1}{|l|}{$\cos
^{2}\phi a_{1}+\sin ^{2}\phi a_{2}$} & \multicolumn{1}{|l|}{$\cos ^{2}\phi
q_{1}+\sin ^{2}\phi q_{2}$} \\ \cline{2-5}
Credit Loan & \multicolumn{1}{|l|}{$d_{1}\sin ^{2}\phi $} & 
\multicolumn{1}{|l|}{$\sin ^{2}\phi y_{1}+\cos ^{2}\phi y_{2}$} & 
\multicolumn{1}{|l|}{$\sin ^{2}\phi a_{1}+\cos ^{2}\phi a_{2}$} & 
\multicolumn{1}{|l|}{$\sin ^{2}\phi q_{1}+\cos ^{2}\phi q_{2}$} \\ \hline
\end{tabular}
\\
\text{Table 1. Usual credit loan compared with quantum credit loan for }M=2%
\text{.}
\end{gather*}%
Should be stressed that the condition $Tr(\overline{A})=Tr(A)=d_{0}$ implies
that $q_{2}=(1+t)\left[ (1+t)d_{0}-q_{1}\right] $, then there is only one
arbitrary parameter $q_{1}$ in a general credit loan with $M=2$ and the
transformed periodic installments can be written as%
\begin{eqnarray}
\overline{q}_{1} &=&(\cos ^{2}\phi -(1+t)\sin ^{2}\phi )q_{1}+(1+t)^{2}\sin
^{2}\phi d_{0}  \label{t2} \\
\overline{q}_{2} &=&(\sin ^{2}\phi -(1+t)\cos ^{2}\phi )q_{1}+(1+t)^{2}\cos
^{2}\phi d_{0}  \notag
\end{eqnarray}%
where now the transformed periodic installments are written in terms of the
initial debt taken by the borrower and the first payment. This
representation of the transformed $\overline{q}_{i}$ is suitable to fix $%
q_{1}=0$, that is, the initial payment of the non-rotated credit loan
vanishes and then $\overline{q}_{1}=\sin ^{2}\phi (1+t)^{2}d_{0}$ and $%
\overline{q}_{2}=\cos ^{2}\phi (1+t)^{2}d_{0}$ which makes explicit the
modulation in $\phi $ introduced by the rotation in the $\overline{q}_{i}$.

Now, let us consider the known loan systems:\ the French and German system.
In the French system, $q_{1}^{(F)}=q_{2}^{(F)}=q^{(F)}=d_{0}\frac{(1+t)^{2}}{%
2+t}$ and then $\overline{q}_{1}^{(F)}=\overline{q}_{2}^{(F)}=q^{(F)}$. This
is expected because in the French system, the periodic installment operator
is diagonal $Q^{(F)}I$, then $UQ^{(F)}IU^{-1}=Q^{(F)}$. In the German
system, $q_{1}^{(G)}=\frac{d_{0}}{2}+td_{0}$ and $q_{2}^{(G)}=\frac{d_{0}}{2}%
+t\frac{d_{0}}{2}$, then%
\begin{eqnarray}
\overline{q}_{1}^{(G)} &=&\left\langle 1\right\vert \overline{Q}\left\vert
1\right\rangle =\frac{d_{0}}{2}+td_{0}\left( 1-\frac{1}{2}\sin ^{2}\phi
\right)   \label{13} \\
\overline{q}_{2}^{(G)} &=&\left\langle 2\right\vert \overline{Q}\left\vert
2\right\rangle =\frac{d_{0}}{2}+td_{0}\left( \frac{1}{2}+\frac{1}{2}\sin
^{2}\phi \right)   \notag
\end{eqnarray}%
Should be noted that, as it happens in the transformed amortization, the sum
of the periodic installments does not depend on $\phi $, that is, $\overline{%
q}_{1}+\overline{q}_{2}=q_{1}+q_{2}$. We can transform the German system
with $q_{1}^{(G)}\neq q_{2}^{(G)}$ into a loan system with identical
periodic installments $\overline{q}_{1}^{(G)}=\overline{q}_{2}^{(G)}$. By
using last equation this condition is obeyed when $\sin \phi =\frac{1}{\sqrt{%
2}}$ which is satisfied for $\phi =\pi /4$. Replacing this value in $%
\overline{q}_{1}^{(G)}$ or $\overline{q}_{2}^{(G)}$ we obtain $\overline{q}%
^{(G)}=\frac{d_{0}}{2}(1+\frac{3}{2}t)$. Then by rotating the basis we can
transform a non-constant periodic payment schedule into a constant\ periodic
payment schedule. Must be stressed that the amortizations in the German
system are constant and remain the same after the $\phi =\pi /4$ rotation.
Then, the new loan system is not exactly a French system but a mixture of
both, that is, the transformed loan have constant periodic installments and
constant amortizations but the periodic installment obtained $\overline{q}%
^{(G)}=\frac{D_{0}}{2}(1+\frac{3}{2}t)$ is not identical to the periodic
installments of the French system $q^{(F)}=d_{0}\frac{(1+t)^{2}}{2+t}$. This
new mixed French-German system credit loan cannot be obtained by the usual
formalism of eq.(\ref{m1}) because if $q_{n}$ and $a_{n}$ are constants then 
$y_{n}$ and $d_{n}$ cannot change. The relation between quantum credit loans
and normal credit loans is analogue to the relation between quantum game
theory and classical game theory, where in the quantum theory, the set of
possible strategies is enlarged by interfering strategies (\cite{pio0} and 
\cite{flit}). The same procedure is introduced with the quantum credit loans
by enlarging the set of repayment schedules by, for example, interfering
normal repayment schedules. In turn, a path integral formulation of credit
loans could be constructed with an appropriate Lagrangian, where the
probability amplitude for a particular path in the debt space (two possible
different paths can be with constant amortization and the other with
constant periodic installment)\ can be calculated with an action $S=\int
L(t)dt$. Interfering paths can be constructed with $e^{iS}$ as it is used in
the Feynman path integral formulation in quantum mechanics or other quantum
model of the stock market (\cite{mont} and \cite{jime}). Finally, an
interesting fact about rewriting the credit loan in a vector space is that
we have the freedom to choose any linear combination of the states of the
loan. There are particular quantum states in physics that are called
coherent states whose dynamics closely resembles the oscillatory behavior of
a classical harmonic oscillator, and is widely familiar in quantum optics
because it describes the maximal coherence and classical behavior of the
quantum electromagnetic field. This is realized as laser light that emit
photons in a highly coherent frequency mode inside a cavity \cite{coh}. In 
\cite{buzek}, the coherent states of $M=2$ is obtained in Appendix (eq.A6),
where a non-trivial phase appears. In the particular case of $M=2$, the mean
values of the loan operators in the coherent state are identical to those
obtained in eq.(\ref{9.1}) but it is expected that for $M>2$, interesting
differences between the rotated and coherent states can be used to improve
the loan properties.

\subsection{$M=3$}

We can go further in the analysis of the basis rotation by considering a
credit loan of duration $M=3$. The orthogonal group $SO(3)$ can be
constructed through its Lie algebra $so(3)$ that contains three generators $%
L_{1}$, $L_{2}$ and $L_{3}$ that corresponds to antisymmetric $3\times 3$
matrices in the adjoint representation \cite{geo}.\footnote{%
In quantum mechanics, these matrices represent the orbital angular momentum
generators that satisfy the commutation relations $\left[ L_{i},L_{j}\right]
=i\epsilon _{ijk}L_{k}$ where $\epsilon _{ijk}$ is the Levi-Civita
completely antisymmetric tensor.} The transformation $U$ can be computed as $%
U=e^{i\theta L_{1}}e^{i\gamma L_{2}}e^{i\phi L_{3}}$. Then, by writing $%
B=\{\left\vert 1\right\rangle ,\left\vert 2\right\rangle ,\left\vert
3\right\rangle \}$, the basis rotation reads%
\begin{gather}
\left\vert \varphi _{1}\right\rangle =\sin \theta \cos \phi \left\vert
1\right\rangle +\cos \theta \cos \phi \left\vert 2\right\rangle -\sin \phi
\left\vert 3\right\rangle  \label{m3.1} \\
\left\vert \varphi _{2}\right\rangle =(\sin \theta \sin \phi \sin \gamma
+\cos \theta \cos \gamma )\left\vert 1\right\rangle +(\cos \theta \sin \phi
\sin \gamma -\sin \theta \cos \gamma )\left\vert 2\right\rangle +\cos \phi
\sin \gamma \left\vert 3\right\rangle  \notag \\
\left\vert \varphi _{3}\right\rangle =(\cos \theta \sin \gamma +\sin \theta
\sin \phi \cos \gamma )\left\vert 1\right\rangle +(\cos \theta \sin \phi
-\sin \theta \sin \gamma )\left\vert 2\right\rangle +\cos \phi \cos \gamma
\left\vert 3\right\rangle  \notag
\end{gather}%
and the similarity transformation $U^{(3)}$ reads%
\begin{equation}
U^{(3)}=\left( 
\begin{array}{ccc}
\sin \theta \cos \phi & \sin \theta \sin \phi \sin \gamma +\cos \theta \cos
\phi \cos \gamma & \sin \theta \sin \phi \cos \gamma +\cos \theta \sin \gamma
\\ 
\cos \theta \cos \phi & \cos \theta \sin \phi \sin \gamma -\sin \theta \cos
\gamma & \cos \theta \sin \phi \cos \gamma -\sin \theta \sin \gamma \\ 
-\sin \phi & \cos \phi \sin \gamma & \cos \phi \cos \gamma%
\end{array}%
\right)  \label{m3.2}
\end{equation}%
where we have three arbitrary angles $\theta ,\phi $ and $\gamma $ to
manipulate freely which defines the parameter space of $SO(3)$. The
transformation of the periodic installment operator $\overline{Q}=UQU^{-1}$
allows us to obtain the mean values $\left\langle n\right\vert \overline{Q}%
\left\vert n\right\rangle $ as%
\begin{gather}
\overline{q}_{1}=\left\langle 1\right\vert \overline{Q}\left\vert
1\right\rangle =\sin ^{2}\theta \cos ^{2}\phi q_{1}+\sec (2\gamma )\cos
^{2}\theta (\cos ^{2}\gamma q_{2}-\sin ^{2}\gamma q_{3})+  \label{m3.3} \\
\sec (2\gamma )\sin ^{2}\theta \sin ^{2}\phi (\cos ^{2}\gamma q_{3}-\sin
^{2}\gamma q_{2})  \notag \\
\overline{q}_{2}=\left\langle 2\right\vert \overline{Q}\left\vert
2\right\rangle =\cos ^{2}\theta \cos ^{2}\phi q_{1}+\sec (2\gamma )\sin
^{2}\theta (\cos ^{2}\gamma q_{2}-\sin ^{2}\gamma q_{3})+  \notag \\
\sec (2\gamma )\cos ^{2}\theta \sin ^{2}\phi (\cos ^{2}\gamma q_{3}-\sin
^{2}\gamma q_{2})  \notag \\
\overline{q}_{3}=\left\langle 3\right\vert \overline{Q}\left\vert
3\right\rangle =\sin ^{2}\phi q_{1}+\sec (2\gamma )\cos ^{2}\phi (\cos
^{2}\gamma q_{3}-\sin ^{2}\gamma q_{2})  \notag
\end{gather}%
Should be noted that when $\gamma =\frac{\pi }{2}$ and $\theta =0$ we
recover a rotation $\phi $ around the $\left\vert 3\right\rangle $ axis. Now
there are three parameters to be tuned in order to satisfy the best periodic
payment schedule for the borrower. By writing $\sin \phi =x$, $\sin \theta
=y $ and $\sin \gamma =z$ last equation can be rewritten as%
\begin{gather}
\overline{q}_{1}=y^{2}(1-x^{2})q_{1}+\frac{1-y^{2}}{1-2z^{2}}\left[
(1-z^{2})q_{2}-z^{2}q_{3}\right] +\frac{x^{2}y^{2}}{1-2z^{2}}\left[
(1-z^{2})q_{3}-z^{2}q_{2}\right]  \label{m3.4} \\
\overline{q}_{2}=(1-x^{2})(1-y^{2})q_{1}+\frac{y^{2}}{1-2z^{2}}\left[
(1-z^{2})q_{2}-z^{2}q_{3}\right] +\frac{x^{2}(1-y^{2})}{1-2z^{2}}\left[
(1-z^{2})q_{3}-z^{2}q_{2}\right]  \notag \\
\overline{q}_{3}=x^{2}q_{1}+\frac{1-x^{2}}{1-2z^{2}}\left[
(1-z^{2})q_{3}-z^{2}q_{2}\right]  \notag
\end{gather}%
These set of equations for the periodic payments $\overline{q}_{i}$ as a
function of $x$, $y$ and $z$ provide a large set of degrees of freedom in
order to manipulate the payment schedule.

\section{Quantum indexed credit loans}

\begin{figure}[tbp]
\centering\includegraphics[width=150mm,height=45mm]{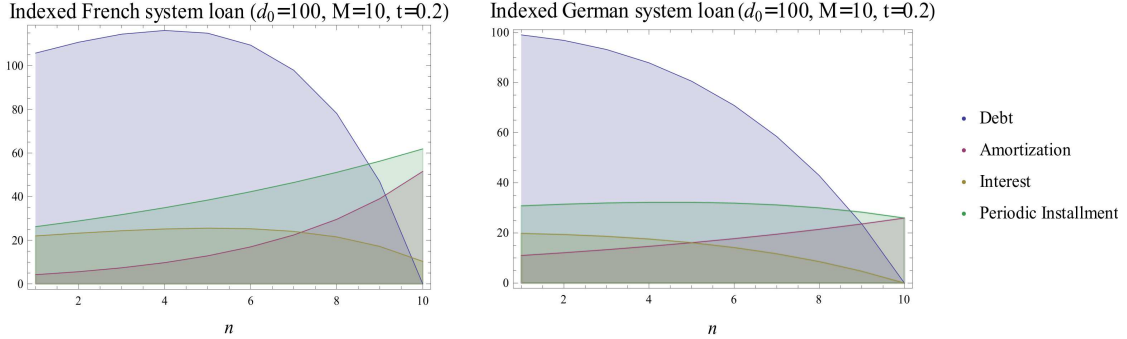}
\caption{Indexed credit loan with $d_{0}=100$, $M=10$, $t=0.2\,$\ and $%
u_{n}=a^{n}$ where $a=1.1$. Left:\ French system. Right:\ German system. }
\label{icl}
\end{figure}

In order to introduce the indexed credit loans in terms of the GHA, we can
recall eq.(\ref{new3.2})%
\begin{equation*}
d_{n}=(1+t_{n-1})d_{n-1}-q_{n}
\end{equation*}%
where in this case the interest rate $t_{n}$ depends on $n$. A particular
example of this loan, is the UVA credit loan in Argentina, where the UVA
(Unidad de Valor Adquisitivo \cite{uva}) is a new variable and the loan
contract is made in units of UVA and not in Argentinian monetary unit
(Argentinian Pesos). The UVA is related to the Argentinian unit currency as $%
u_{n}=\alpha \times 1$peso, where $\alpha $ is a number that varies from day
to day and is computed by using the Consumer's price index (CPI). The
procedure to create this relation between UVA and CPI is by indexing $\alpha 
$ to the Reference Stability Coefficient Rate (CER) coefficient \cite{cer}
which is elaborated by the Central Bank and depends on the Argentinian
inflation.\footnote{%
The stabilized reference coefficient (CER) is computed daily by using the
three previous Consumer's price index (CPI) by the formula $f_{t}=\left( 
\frac{IPC_{j-2}}{IPC_{j-3}}\right) ^{l/k}$ for the initial six days of the
months and $f_{t}=\left( \frac{IPC_{j-1}}{IPC_{j-2}}\right) ^{l/k}$ for the
remaining days of the months, where $j$ is the current month, $k$ is the
number of days corresponding to the current month and $f_{t}$ is the daily
actual factor. The CER\ coefficient is computed as $CER_{t}=f_{t}\times
CERt_{t-1}$. In this sense, the CER coefficient is more stable than the
inflation.} That is, the units of UVA\ depend on the Inflation or Consumer's
Index Price. The UVA\ credit loan is a French system loan in the UVA\ units
but the borrower makes the payment in Argentinian monetary unit. Then the
amount to be paid periodically is $q_{n}=u_{n}q$ where $q$ is the periodic
payment in UVA units obtained from the French system and $u_{n}$ is the
price of the UVA unit in term of the Argentinian monetary unit. Then,
although the periodic payment is in UVA units and is constant in time, the
amount to be paid with Argentinian monetary unit varies with time because
the UVA price in terms of Argentinian pesos depends on time. Going further,
although the debt in UVA unit follows the same behavior as the debt in the
non-indexed French system, the debt in Argentinian monetary unit for the
indexed loan follows a different behavior. That is, although the debt $d_{n}$
in UVA\ unit decreases, the debt in Argentinian monetary unit is variable.
Computing the amount of debt in Pesos as%
\begin{equation}
\overline{d}_{n}=d_{n}u_{n}  \label{il1}
\end{equation}%
Then it can be shown that $\overline{d}_{n}$ follows the recurrence relation%
\begin{equation}
\overline{d}_{n}=(1+t)\frac{u_{n}}{u_{n-1}}\overline{d}_{n-1}-u_{n}q
\label{il2}
\end{equation}%
which implies an effective indexed interest rate $t_{n-1}=(1+t)\frac{u_{n}}{%
u_{n-1}}-1$. If $\frac{u_{n}}{u_{n-1}}=1$, $t_{n-1}=t$ as it is expected.
From eq.(\ref{il3}), when $u_{n}$ increases with $n$ and $d_{n}$ decreases,
there is a maximum value that can be obtained through the equation $\frac{d}{%
dn}(\overline{d}_{n})=\frac{d}{dn}(d_{n})u_{n}+d_{n}\frac{du_{n}}{dn}=0$ for
a particular value $n$. In figure \ref{icl} the behavior of the debt,
interest, amortization and periodic payment is shown as a function of $n$
when the UVA\ credit loan follows the French system (left) and when follows
the German system (right). In the figure \ref{icl} it can be seen that the
debt $\overline{d}_{n}$ in the\ German system does not present a peak
because the schedule payment $q_{n}^{(G)}$ decreases with $n$ and balance
the increment due to $u_{n}$.\footnote{%
The debt and periodic payment in the indexed UVA system using the German
system \ for the UVA\ units is more suitable for the borrower than the
indexed UVA system using the French system for the UVA\ units.} 

\begin{figure}[tbp]
\centering\includegraphics[width=100mm,height=60mm]{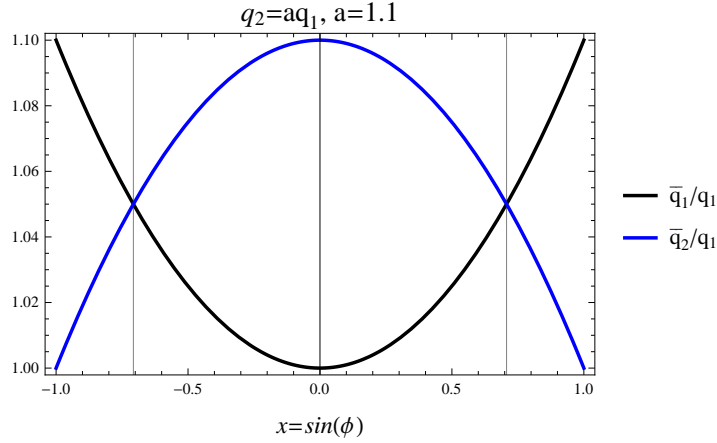}
\caption{Transformed periodic installments $\overline{q}_{1}/q_{1}$ and $%
\overline{q}_{2}/q_{1}$ as a function of $x=\sin \protect\phi $ with $a=1.1$%
. Vertical lines indicates the values $x=\pm 1/\protect\sqrt{2}$. }
\label{a1}
\end{figure}

\begin{figure}[tbp]
\centering\includegraphics[width=150mm,height=70mm]{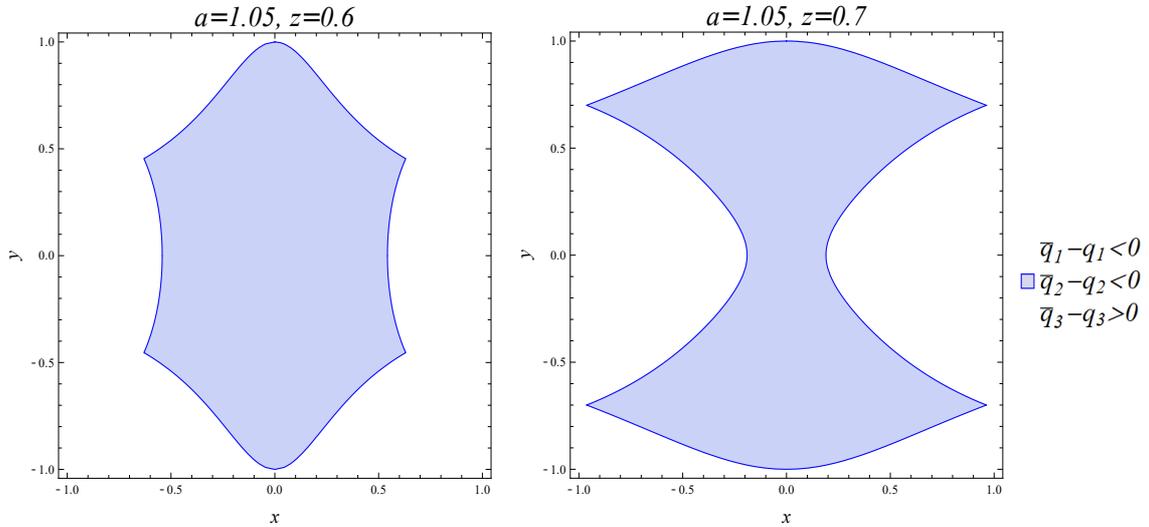}
\caption{Regions where $\overline{q}_{1}-q_{1}<0$, $\overline{q}_{2}-q_{2}<0$
and $\overline{q}_{3}-q_{3}>0$ holds, with $a=1.05$. Right: $z=\sin \protect%
\gamma =0.6$. Left:\ $z=\sin \protect\gamma =0.7$. \ }
\label{region}
\end{figure}

To introduce the indexed UVA credit loans in terms of GHA, we can recall the
function $f(D)$ of eq.(\ref{new1}) that is replaced by $f(D)=(T-I)D$ where $%
T $ is the interest rate operator that satisfies $\left[ T,D\right] =0$, $I$
is the identity operator and $T\left\vert n\right\rangle =t_{n}\left\vert
n\right\rangle $ where $t_{n}=(1+t)\frac{u_{n+1}}{u_{n}}-1$ in the
particular case of the UVA credit loan. For simplicity we can consider $M=2$
and the transformed periodic payments read%
\begin{eqnarray}
\overline{q}_{1} &=&\cos ^{2}\phi u_{1}q+\sin ^{2}\phi u_{2}q  \label{il3} \\
\overline{q}_{2} &=&\sin ^{2}\phi u_{1}q+\cos ^{2}\phi u_{2}q  \notag
\end{eqnarray}%
For simplicity we can write $u_{2}=au_{1}$ where $a$ is the inflation rate
between the two periods of the loan and we obtain%
\begin{eqnarray}
\frac{\overline{q}_{1}}{u_{1}q} &=&\cos ^{2}\phi +a\sin ^{2}\phi  \label{il4}
\\
\frac{\overline{q}_{2}}{u_{1}q} &=&\sin ^{2}\phi +a\cos ^{2}\phi  \notag
\end{eqnarray}%
then we can find that $\overline{q}_{1}=\overline{q}_{2}$ for $\phi =\pi /4$
and we have transformed the indexed credit loan into a constant periodic
installment loan (see figure \ref{a1}), where this constant is $\overline{q}%
_{1}(\phi =\frac{\pi }{4})=\overline{q}_{2}(\phi =\frac{\pi }{4})=\frac{1}{2}%
(1+a)u_{1}q$. 
Then we obtain 
\begin{equation}
\overline{q}_{1}(\phi =\frac{\pi }{4})=(\frac{1}{2}+\frac{a}{2})q_{1}\text{
\ \ \ \ \ \ \ \ \ }\overline{q}_{2}(\phi =\frac{\pi }{4})=(\frac{1}{2}+\frac{%
1}{2a})q_{2}  \label{il5}
\end{equation}%
which in the case $a>1$, $\overline{q}_{1}>q_{1}$ but $\overline{q}%
_{2}<q_{2} $ as it was shown in the last section. That is, we compensate the
extra amount paid in the first period by paying $\overline{q}_{2}<q_{2}$.
For $M=2$ the angle of the rotation $\phi $ does not depend on $a$. By
considering $M=3 $ and considering a constant inflation rate $a$, then $%
u_{2}=au_{1}$ and $u_{3}=au_{2}=a^{2}u_{1}$, then we can compute the
conditions for $x$, $y$ and $z$ in order to have the possible configuration $%
\overline{q}_{1}-q_{1}<0 $, $\overline{q}_{2}-q_{2}<0$ and $\overline{q}%
_{3}-q_{3}>0$ (figure \ref{region}). The shaded region with the specific
values $z=0.6$ and $z=0.7$ indicates possible values for $x$ and $y$ that
allows the configuration to be achieved. The different schedules can be
given by noting that $\overline{q}_{1}-q_{1}+\overline{q}_{2}-q_{2}+%
\overline{q}_{3}-q_{3}=0$, then in the case $\overline{q}_{1}-q_{1}<0$, then 
$q_{2}-\overline{q}_{2}<\overline{q}_{3}-q_{3}$, which implies that the
transformed schedule payment contains at least one transformed periodic
payment that is larger than the non-transformed one. 

An interesting point to be discussed is that the dependence of $u_{n}$ with $%
n$ cannot be known because the inflation rate is actualized periodically and
this value depends on several factors, such as the monetary basis, dollar
price, etc. The similarity transformation $U$ implies that the transformed
periodic payments are a superposition of the original periodic payments, but
these cannot be known at the beginning of the loan because of the lack of
knowledge of $u_{n}$. If for simplicity we consider again $M=3$ and suppose
that the first payment cannot be addressed by the borrower, a rotation is
still available that gives eq.(\ref{il3}) where $u_{1}$ is known but $u_{2}$
and $u_{3}$ are not known. The customer needs to perform a rotation in such
a way to obtain $\overline{q}_{1}<q_{1}=u_{1}q$. At this point it could be
necessary to consider the future inflation expectations in order, at least,
to define a probability density function $p(u_{2},u_{3})$ that obeys $%
\int_{0}^{u_{\max }}\int_{0}^{u_{\max }}du_{2}du_{3}p(u_{2},u_{3})=1$, in
such a way to determine the best option for the customer. By using eq.(\ref%
{m3.4}), it can be seen that $\overline{q}_{i}=\overline{q}%
_{i}(q_{1},x,y,z,u_{1},u_{2},u_{3})$, then we can compute $\left\langle 
\overline{Q}_{i}\right\rangle =\int_{0}^{u_{\max }}\int_{0}^{u_{\max
}}du_{2}du_{3}p(u_{2},u_{3})\overline{Q}_{i}(q_{1},x,y,z,u_{1},u_{2},u_{3})$%
, where $u_{\max }$ is an upper bound for the indexed variable and then
consider the most suitable election of $x$, $y$ and $z$ as a function of $%
u_{\max }$ for the borrower. Another solution is to model the future
expectation inflation by some particular function $u_{n}$. In the
Argentinian case, by fitting the real values of $u_{n}$ \cite{uva} there are
two possible reliable solutions: a power law fitting $u_{n}=u_{0}e^{\alpha
n} $ where $\alpha \sim 0.0109$ and $u_{0}=14.27$ and a linear fitting $%
u_{n}=u_{0}+0.03n$. Any of these two fittings allow to rotate the basis of
the Hilbert space of the loan in a specific way to obtain, for example,
constant periodic payments, giving predictability to the borrower.

The general procedure for large $M$, such as mortgage loans, that can be
have a duration of $20$ or $30$ years, implies to develop complex strategies
in order to take advantage of the large matrix transformation $U$ and the
different payment schedules that can be obtained from the equation $Tr(%
\overline{Q})=Tr(Q)$. These strategies involve to use the mathematical
machinery of group theory in order to use the subgroups of $SO(M)$ to
introduce selective payment schedules and can contribute to the analysis of
quantum finance interest rate models (\cite{baq}, \cite{baq2}). For example,
in the case $M=3$ we can consider a rotation subgroup as%
\begin{equation}
U=\left( 
\begin{array}{ccc}
\cos \phi  & \sin \phi  & 0 \\ 
-\sin \phi  & \cos \phi  & 0 \\ 
0 & 0 & 1%
\end{array}%
\right)   \label{il6}
\end{equation}%
which implies not altering the $q_{3}$ payment. This is a particular case of 
$SO(2)\subset SO(3)$. A large classification of subgroups of $SO(M)$ can be
found in the literature in terms of cosets, conjugacy classes, etc (see \cite%
{geo} chapter 19 to 25). From the perspective of the Lie algebra $so(M)$,
the subgroups can be found by exponential map of the specific generators of
the rotation. In turn, using the Lie algebra it is possible to obtain, for
example in the case $M=4$ that $so(4)=so(3)\oplus so(3)$. By computing the
exponential map of the generators of each $so(3)$ it is possible to obtain
specific subgroup of rotations according to the customer needs. This
procedure can be generalized for large $M$ without complications, although
implementing a software that allows to manipulate the entire parameter space
of the Lie algebra and to select specific subgroups of $SO(M)$ through the
Lie algebras should be extremely necessary. To make a final remark, the
generalized Heisenberg algebra is suitable to be applied to quantum finance
related to index-linked coupons that depends on market defined index \cite%
{baq3}. In turn, diverse Hamiltonians are defined to model the stock market.
For example, in \cite{tgao} in which a quantum anharmonic oscillator is used
as a model for the stock market. The motion of the stock price is modeled as
the dynamics of a quantum particle and the probability distribution of price
return is computed with the wave function obtained from the Scrhodinger
equation with an anharmonic oscillator potential. The Black-Scholes equation
can be mapped into the Schrodinger equation (see \cite{cont} and \cite{have}%
) and a Hamiltonian operator with potential can be constructed. Interesting
approaches using non-Hermitian Hamiltonians are given in \cite{baqq}. In 
\cite{cotf} a finite dimensional Hilbert space is constructed to model the
stock market that isomorphic to $%
\mathbb{C}
^{d}$, where $d$ is the discrete number of possible rates of return. In \cite%
{schm} all possible realizations of investors holding securities and cash is
taken as the basis of the Hilbert space of market states. The temporal
evolution of an isolated market is unitary in this space. In \cite{orr}, the
quantum framework is used to model supply and demand. The different
potentials appearing in the Hamiltonians of the different models of the
stock market can be studied using the generalized Heisenberg algebras (see 
\cite{berr}). Then, having access to the algebra of observables of a quantum
system allows us to design better general quantum models for stock price\ or
financial instruments instead of using a particular representation of the
wavefunction (the coordinate representation) that is not more than a
particular representation of the underlying algebra.

Finally, should be stressed that the commutative operators chosen to define
the loan systems (eq.(\ref{new1})a) is suitable to generalize via
non-commutative operators, which can be used to capture the impossibility of
a joint measurement of these quantities and hence can explain the existence
of order effects and deeper levels of ambiguity in respect to the loan
realization. In turn, the quantum formalism introduced for credits is
suitable to consider repackaging of different types of loans such as credit
loans, credit card debt, business loans and mortgage into pools, because
these pools can be modeled as tensor product of the algebras. The symmetry
group of the tensor product of the algebras will allows to construct
entangled loan states, without changing the return because $Tr(\overline{Q}%
_{1}\otimes \overline{Q}_{2}\otimes ...\otimes \overline{Q}%
_{n})=Tr(Q_{1}\otimes Q_{2}\otimes ...\otimes Q_{n})$.

\section{Conclusions}

In this work we have shown a generalization of a credit loan by introducing
linear operators for the debt, amortization, interest and periodic
installments that act on a vector space of dimension $M$, where $M$ is the
loan duration. We have shown that endowing this vector space with an $SO(M)$
symmetry, a basis rotation of the orthonormal basis allows us to change the
schedule of the periodic payments, allowing better benefits for the borrower
in the case a payment cannot be afford or any other circumstance in which a
default of the remaining debt is possible. The values of the loan operators
are computed as mean values of these operators in any of the orthogonal
vectors of the eigenbasis of the vector space. In turn, the rotation basis
does not change the total amortization and the total amount paid by the
borrower, which is expected in order not to change what the lender earns and
simultaneously not to change what the borrower amortizes. In turn, it was
shown that a generalized Heisenberg algebra for the loan operators can be
defined that gives the usual recurrence relations for the debt,
amortization, interest and periodic installments. Finally, the case of
indexed credit loans is studied, where the rotation basis provides a
sophisticated solution when the periodic installments depend on exogenous
variables with a non-predictive behavior. Some particular cases with $M=2$
and $M=3$ are studied showing how the rotation transformation can be applied
to design specific schedule periodic payments.

\section{Acknowledgments}

This paper was partially supported by grants of CONICET (Argentina National
Research Council) and Universidad Nacional del Sur (UNS) and by ANPCyT
through PICT 1770, and PIP-CONICET Nos. 114-200901-00272 and
114-200901-00068 research grants, as well as by SGCyT-UNS., J. S. A. is a
member of CONICET.

\end{document}